\documentclass{elsart}

\usepackage{amssymb}

\usepackage{amsmath}
\usepackage{amssymb}

\usepackage{url}

\usepackage[sort,numbers]{natbib}  

\usepackage{verbatim}  

\usepackage{graphicx}
\newcommand{\nsixteen}{\ensuremath{\text{${}^{16}$N}}}
\newcommand{\lieight}{\ensuremath{\text{${}^8$Li}}}

\newcommand{\boroneight}{\ensuremath{\text{${}^{8}$B}}}

\newcommand{\sz}{\scriptsize}

\newcommand{\bi}{\begin{itemize}}
\newcommand{\be}{\begin{enumerate}}
\newcommand{\ei}{\end{itemize}}
\newcommand{\ee}{\end{enumerate}}
\newcommand{\heavywater}{{\ensuremath{\text{D$_2$O}}}}

\newcommand{\lightwater}{{\ensuremath{\text{H$_2$O}}}}

\newcommand{\dg}{\ensuremath{^{\circ}}}  
\begin{document}

\begin{frontmatter}



\title{Optical calibration hardware for the {S}udbury {N}eutrino {O}bservatory}


\author{B.A. Moffat},
\author{R.J. Ford},
\author{F.A. Duncan},
\author{K. Graham},
\author{A.L. Hallin},
\author{C.A.W. Hearns},
\author{J. Maneira\corauthref{cor:author}\thanksref{foot:lip}},
\ead{maneira@lip.pt}
\author{P. Skensved}
\corauth[cor:author]{Corresponding author.}
\thanks[foot:lip]{Present address: La\-bo\-ra\-t\'{o}\-rio de
Ins\-tru\-menta\-\c{c}\~{a}o e F\'{\i}\-si\-ca Ex\-pe\-ri\-men\-tal de
Par\-t\'{\i}\-cu\-las (LIP), Av. Elias Garcia, 14,
1$^{\circ}$, 1000-149 Lisboa, Portugal.}
\address{Queen's University, Physics Department, Kingston, Ontario, Canada K7L 3N6}

\author{D.R. Grant\thanksref{foot:cwru}}
\address{Carleton University, Physics Department, Ottawa, Ontario, Canada K1S 5B6}
\thanks[foot:cwru]{Present address: Department of Physics, Case Western Reserve 
University, Cleveland, OH 44106, USA.}

\begin{abstract}


The optical properties of the Sudbury Neutrino Observatory (SNO)
heavy water Cherenkov neutrino detector are measured \emph{in situ} using
a light diffusing sphere (``laserball''). This diffuser is connected to a pulsed
nitrogen/dye laser via specially developed underwater optical fibre
umbilical cables.
The umbilical cables are designed to have a small bending radius, and can be 
easily adapted for a variety of calibration sources in SNO.
The laserball is remotely manipulated to many positions in the \heavywater\ and
\lightwater\ volumes, where data at six 
different wavelengths are acquired.  These data are
analysed to determine the absorption and scattering of light in the
heavy water and light water, and the angular dependence of the response of the detector's
photomultiplier tubes.
This paper gives details of the physical properties, construction,
and optical characteristics of the laserball and its associated
hardware.
\end{abstract}

\end{frontmatter}
\section{Introduction}
\label{sect:introduction}

The Sudbury Neutrino Observatory (SNO) detector is designed to detect Cherenkov light
produced by solar neutrino interactions in heavy water (\heavywater).
The accuracy of the solar neutrino measurements depends on a detailed
knowledge of the detector operating conditions, of which the optical
properties play a dominant role.

The SNO detector distinguishes itself from other large-volume water Cherenkov 
detectors in its use of 1\,kt of \heavywater\ contained in a 12\,m diameter 
acrylic vessel (AV) \citep{sno:00}.
The AV is suspended in \lightwater\ which provides shielding from natural
radioactivity in  the detector components and the surrounding rock cavity.
To provide shielding from cosmic rays, the detector is located 2\,km 
underground in the INCO Creighton mine near Sudbury, Canada.
Light emitted in the detector is captured by 9456 photomultiplier tubes (PMTs)
arranged on a 17\,m diameter geodesic support structure (PSUP).
The determination of light attenuation in
the heavy and light water and the PMT response as a function of incident angle 
are part of the optical calibration (OCA) \citep{Moffat:01,Ford:98,Ford:93,Grant:04}.
The determination of the relative timing offsets of the PMT electronic
channels are part of the PMT calibration (PCA).  These calibrations determine
crucial parameters for the precise reconstruction of the energy, position
and direction of neutrino candidate events used in the analyses of solar
neutrinos
\citep{snofirst:01,snonc:02,snodaynight:02,snosaltprl:03,snonsp:05,snoperiod:05},
as well as for the searches for nucleon decay via  'invisible' modes
\citep{snonuc:04}, electron antineutrinos \citep{snoanti:04}, supernova
neutrinos \citep{Heise:01} and analyses of atmospheric neutrinos and muons
\citep{Tagg:01}.
 
\begin{figure}
    \begin{center}
      \includegraphics[width=5.4in]{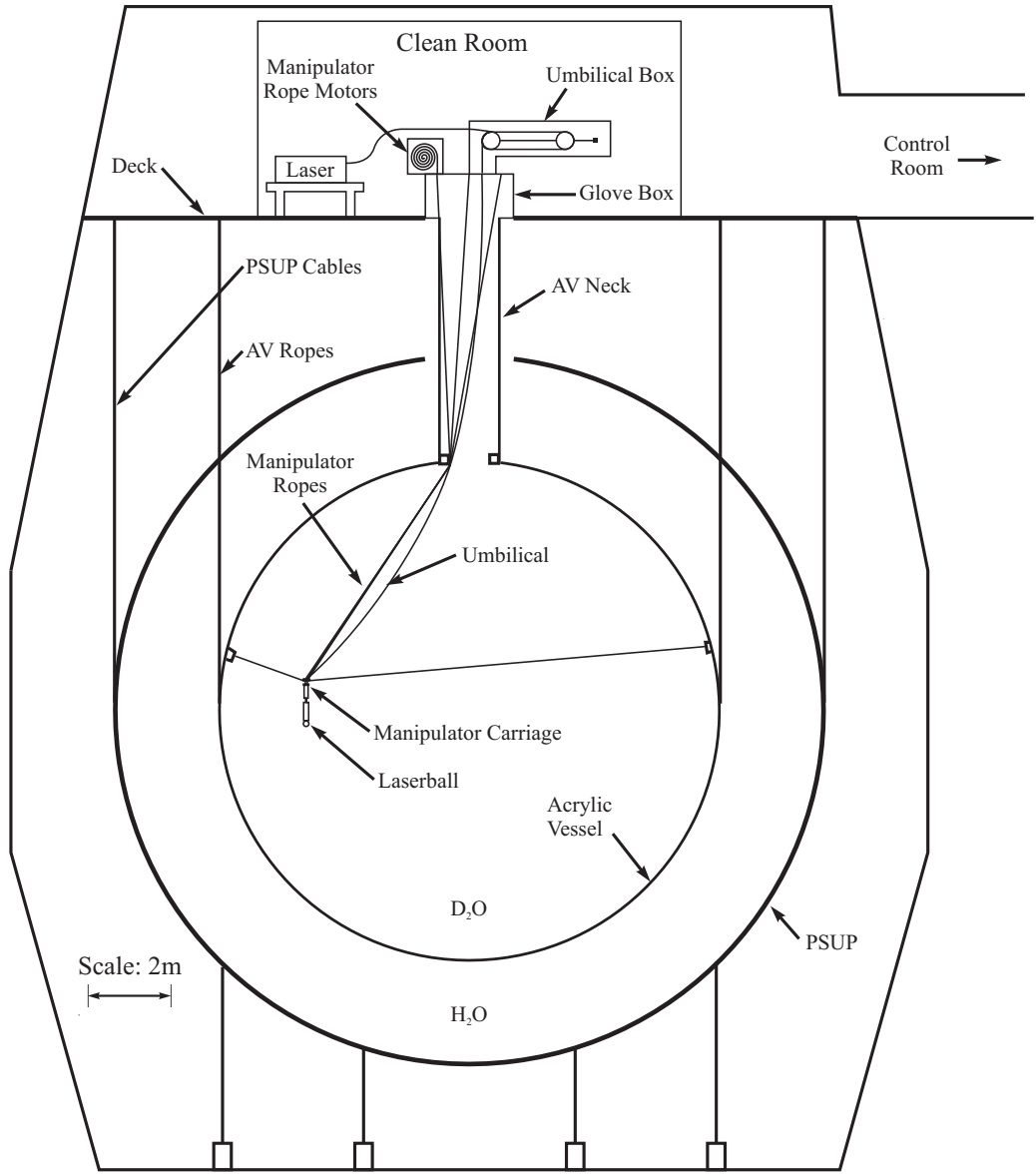}
    \end{center}
    \caption{Overview of the SNO calibration source deployment system. The laserball
	and its associated umbilical cable are also shown.}
    \label{fig:laserball_system}
\end{figure}

The optical calibration hardware described in this paper consists of a light
diffusing sphere (``laserball''), an underwater optical fibre umbilical cable and
a pulsed dye-laser light source. These components are shown in 
Figure~\ref{fig:laserball_system}, along with the source manipulator system that is 
used to manoeuvre different calibration sources, including the laserball, inside the AV.
An overview of the SNO detector design, including a description of the manipulator 
system, is given in Ref.~\citep{sno:00}.
The overall system requirements are briefly described in the next section.
The following three sections contain a detailed description of the design
and construction of each of the optical calibration subsystems, followed by a brief 
summary of the laserball performance in SNO.

\section{System requirements}
\label{sect:requirements}

The OCA hardware must provide good coverage of the detectable
Cherenkov spectrum, which extends from 300 to 700\,nm. The 
pulse intensity must be adjustable for single photoelectron level,
 at $\sim 5\%$ of PMTs triggered per laser pulse, equivalent to a ratio of multi-
 to single-photon hits of $\sim 2.5\%$, in order to maximize the
timing measurement precision. 

While the OCA uses data taken at many laserball positions, the PCA 
is done with the laserball 
only at the centre of the detector and at 505\,nm only. 
The choice of this wavelength for the PCA is done for several reasons:
\begin{itemize}
\item minimum effects from scattering (Rayleigh scattering 
coefficient is about $1.8 \times 10^{-5}$\,cm$^{-1}$ at 505\,nm);
\item reduced time-delayed fluorescence in the optical fibres (low level below
400\,nm, reaching $13 \pm 4\%$ at 337\,nm \citep{Ford:98}, below detection above
400\,nm);
\item high PMT quantum efficiency (85\% of the peak value, which is $\sim
21.5\%$ at 440 nm);
\item good transmission in all the materials: acrylic 
($>90\%$ above 370\,nm), \lightwater\ ($>90\%$ between 280 and 540\,nm) and 
\heavywater\ ($>90\%$ above 340\,nm).
\end{itemize}

Good timing of the detected photons is required by both the OCA and the PCA 
to allow the discrimination between light
coming directly from the source and late light which has been scattered
or reflected from other detector elements \citep{Moffat:01}.
Since the PMT hit times are used in the event reconstruction,
the PCA requires the optical pulse timing width -- determined essentially by the laser pulse
width and by the dispersion in the fibres and the laserball -- to be small 
relative to the intrinsic 1.5\,ns resolution of the PMTs \citep{sno:00}.

\section{Laserball}
\begin{figure}
  \begin{center}
    \includegraphics[width=5in]{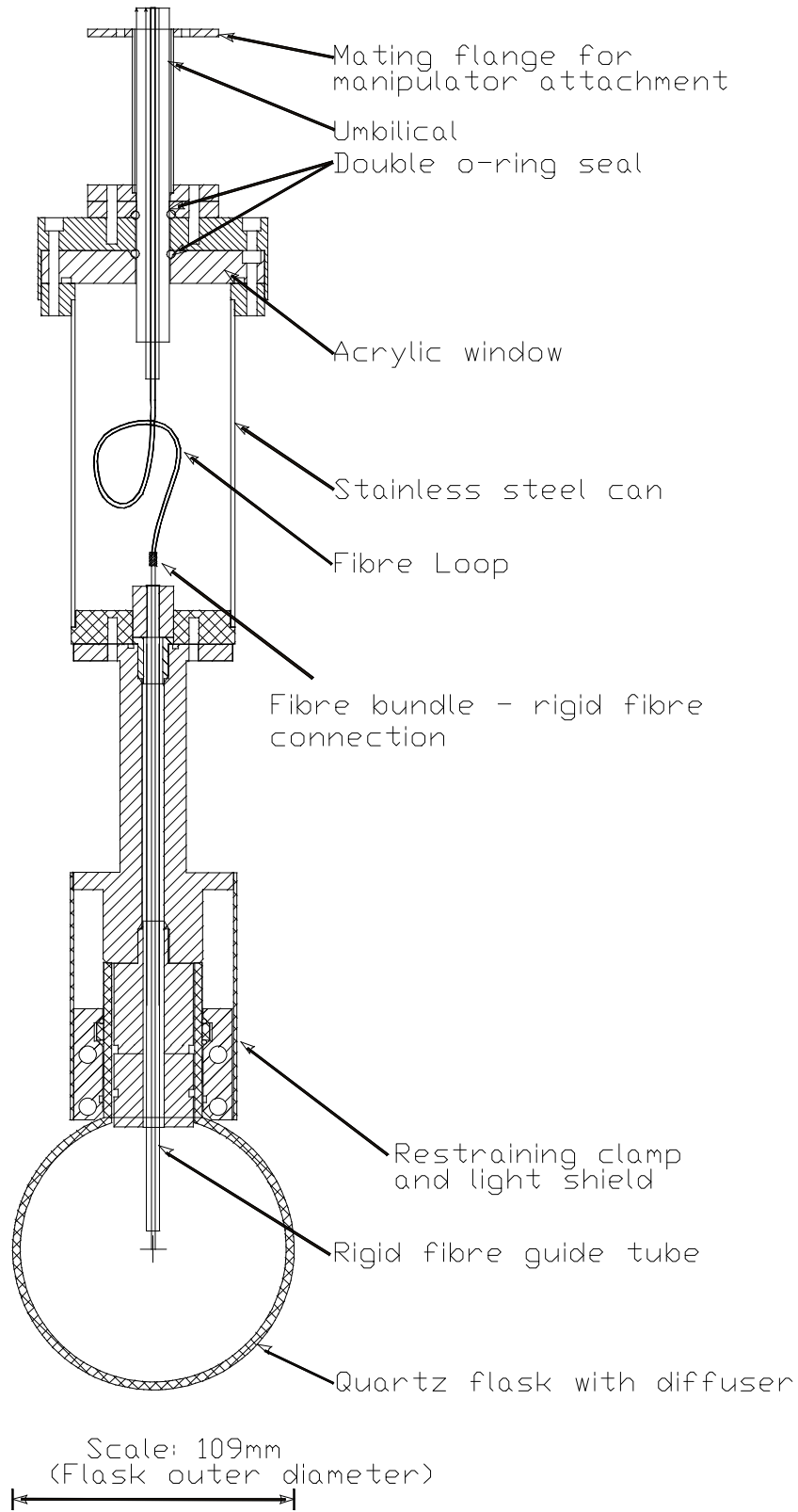}
  \end{center}
  \caption{Cross-sectional view of the laserball.}
  \label{fig:newlaserball}
\end{figure}

The laserball was designed and built to calibrate the optical characteristics
 of the SNO detector
by emitting a quasi-isotropic light distribution in the far-field ({\em i.e.},
at distances large compared to the laserball's dimensions)
\citep{Ford:93,Ford:98,Moffat:01,Grant:04}.  
The final design of the laserball, described in this paper and shown schematically in
Figure~\ref{fig:newlaserball}, was achieved after a series of iterations.

The laserball consists of a 10.9\,cm diameter quartz flask filled with 
0.5\,kg of silicone gel \citep{gesilicones:rtv}, in which 2\,g of 50\,$\mu$m 
diameter air-filled hollow glass spheres \citep{3m:spheres} are suspended.
In the assembly of the laserball, uniform distribution of the
scattering spheres is achieved by continuous agitation
of the flask during the 15~minutes cure time of the silicone gel. It was verified
that bubbles formed in the gel (so it had to be refilled) if the laserball was 
brought to the surface from the underground SNO laboratory.

The mounting hardware is made of stainless steel except for the acrylic
window (for inspection of the fibre loop during assembly).  
The quartz flask has a
straight neck with a restraining ring, and is completely filled with silicone
gel.

The polished tip of the fibre bundle coming from the umbilical cable (see 
Section \ref{sec:umbilical})
is coupled to a single rigid fibre, the bottom tip of which is placed slightly 
above the
centre of the laserball such that the light emitted by the fibres is
redirected by refraction and total internal reflection at the glass-air
interface inside the hollow glass spheres.  This results in a smoothly
varying, quasi-isotropic far-field intensity distribution.  This 
distribution is only weakly dependent on the wavelength even for low
concentrations of scatterers.
The mean free path of 
light between the hollow glass spheres is $\sim 1$\,cm, which is a compromise 
between guaranteeing far-field uniformity and reducing time dispersion and 
light absorption. Monte Carlo simulations of the laserball
\citep{Ford:93} indicate that the light appears to come from a diffuse region with
an effective diameter of 4\,cm.

The far-field light distribution of the
laserball can be adjusted by moving the fibre tip: a displacement of 1\,mm
results in a change of approximately 10\% in the intensity pattern at
385\,nm.  Only a coarse adjustment is required, since the PCA depends only on
the timing, and for the OCA the laserball light distribution is fit along with the
optical parameters using an optical model of the detector
\citep{Moffat:01}
(as it will be described in Section~\ref{sect:performance}).

In addition to the small residual anisotropy in the light distribution
of the laserball due to the placement of the optical fibres, there is a 
larger asymmetry due to shadowing by the mounting hardware.  The light within 
approximately
60\dg\ of the vertical is progressively shadowed, reaching a reduction of
$\sim 50\%$ directly above the laserball.  To minimize reflections from the
mounting hardware and to prevent the detection of light emitted from the neck
of the flask, a cylindrical polished stainless steel light shield extends to
just above the ball surface.

\section{Umbilical cables\label{sec:umbilical}}

The multi-purpose underwater ``umbilical'' cables were designed to be flexible
and robust for use with the source manipulator system, and to provide services
for the operation of all the SNO calibration sources \citep{Moffat:01}. The
optical fibre umbilical contains a bundle of optical fibres for transmitting light from
the laser down to the laserball. This section describes the requirements, design
and fabrication process of the umbilical cables.

\begin{figure}
  \begin{center}
    \includegraphics{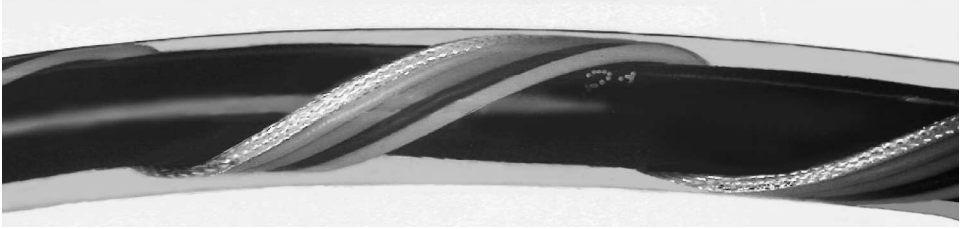}
  \end{center}
  \caption{Umbilical cable section.  The outer diameter of the umbilical is 12.7\,mm.
  The twisted wires correspond to the thin coaxial and hook-up wires.}
  \label{fig:umbilical_section}
\end{figure}

\subsection{Requirements and design}

Design requirements for the umbilicals included:
\begin{itemize}
\item Radioactivity cleanliness: suitable for deployment in SNO for periods of up to
  several weeks at a time without introducing a significant $^{222}$Rn contamination 
  (emanation lower than 10~mBq, which correspond to 10\% of the level in the 
  \heavywater);
\item Impermeability to water to protect cable components and the source itself;
\item High flexibility with a small bending radius, for use with
the manipulator;
\item Length of 30\,m, to reach the farthest available positions inside SNO;
\item Neutral buoyancy in \heavywater\ with constant linear density over
  its length.
\end{itemize}

\begin{figure}
  \begin{center}
    \includegraphics{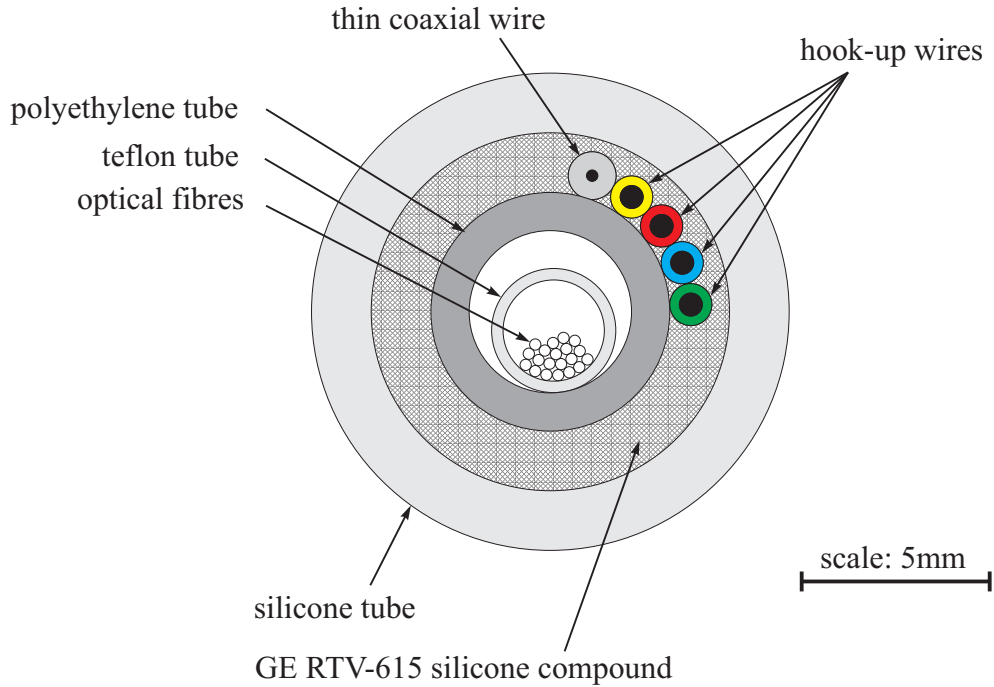}
  \end{center}
  \caption{Cross-sectional view of the laserball umbilical.}
  \label{fig:umbcross}
\end{figure}

Cables with these specifications were not available commercially,
so the expertise to construct them was developed within the SNO Collaboration.
A short length of umbilical cable is shown in Figure~\ref{fig:umbilical_section},
with its cross-sectional view depicted in Figure~\ref{fig:umbcross}.

An umbilical consists of a 30.5\,m long silicone tube with an outer diameter (OD)
of 12.7\,mm containing a smaller 6.35\,mm OD polyethylene tube \citep{mcmastercarr:tubing}
helically wrapped with four thin Teflon-insulated hook-up wires and a thin coaxial
cable \citep{belden:wires}.
The four wires and the coaxial cable centre the polyethylene tube inside the
umbilical, and the helical wrapping permits longitudinal tensile and
compressive forces on the wires to be relieved locally as the umbilical goes
over the pulleys of the manipulator system.
Both of these effects are important in reducing the minimum bending radius
for use with the manipulator system with its 15.2\,cm and 20.3\,cm diameter pulleys.
Additionally, the central polyethylene tube provides mechanical support to the fibres, both
radially (against crushing) as well as longitudinally (against stretching).

For the laserball umbilical, 20~optical fibres \citep{fiberguide:fibres} were
used to allow for redundancy in the case of breakage, and to simplify the
injection of light from the laser while preserving adequate light intensity
at the laserball.
The high-OH (hydroxyl) core fibres provide low attenuation and low time
dispersion over the wavelength range 337--619\,nm probed in the optical
calibration.

The optical fibres are put in a small Teflon tube
\citep{mcmastercarr:tubing}, which protects them during assembly and
deployment.  The fibres are deliberately kept loose inside the Teflon tube so
that they do not take any load and can move relative to the umbilical.  Having
the fibres on the neutral bending axis of the umbilical also greatly reduces
mechanical stresses during deployment.

Variations on the basic umbilical design include replacing the fibre bundle
with a gas capillary for two short-lived radioactive gas sources (\nsixteen\ 
\citep{Drag:01} and \lieight~\citep{Tagg:02}),
replacing the 6.35\,mm polyethylene tube with a high-voltage cable \citep{belden:wires}
for a small proton-tritium accelerator source \citep{Poon:00}, or a fast coaxial 
cable \citep{timesmicrowave:fastcoax} for a thoriated anode
proportional counter source \citep{sno:00}.  Thus one basic umbilical design has been adapted to
provide services to all the SNO calibration sources.

\subsection{Umbilical Fabrication}
The umbilicals are made by first wrapping the four small hook-up wires and the 
thin coaxial wire around the polyethylene tube, and this assembly is then drawn 
into the silicone tube. 
The volume between the silicone tube and the polyethylene tube is filled with
a clear, transparent, two-component liquid silicone \citep{gesilicones:rtv} which
forms a robust, impermeable layer between the silicone tube and the inner
components.
The filling of the umbilicals with silicone is a critical step in their
manufacturing.  At room temperature, the highly viscous ($\sim
4000$\,Pa$\cdot$s) RTV615 silicone remains fluid for four to six hours after
its components are mixed.  During this time, the viscosity increases steadily until the
fluid stops flowing freely.  Therefore, a high-pressure injection system was
developed to force the RTV615 into the umbilical within the first hour after
mixing.

To hold the silicone tubing in place, it is inserted into eight 3.65\,m and one 
1.3\,m sections (total length: 30.5\,m) of thin-walled 15.875\,mm OD copper 
piping.
The inner diameter of the copper piping is only slightly larger than the OD of
the silicone tube.  The silicone tube is wrapped back around the ends of the
copper pipe to form a seal and is ``inflated'' by evacuating the volume
between the copper pipes and the outside of the silicone tubing.  This anchors
the silicone tube to the copper pipe by friction and allows the
helically-wrapped polyethylene tube to be drawn through the silicone tube
without stretching the latter.

The umbilical in its copper pipe housing is connected at one end to an
injection piston and at the other to an evacuated drum, which eliminates
trapped air between cables or at the front edge of the flow of liquid
silicone.  At the injection end of the umbilical, a special ``T''-fitting is
used to ensure the silicone is forced into the annular region between
the polyethylene and the silicone tubes.  The injection piston was designed to
fulfill two goals: degassing the silicone immediately after mixing
at partial vacuum pressure ($\lesssim 3.3$\,kPa),
and then injecting the silicone into the umbilical at
relatively high pressure ($\sim 2.1$\,MPa).  Because of the limited time
available for the umbilical filling, the degassing step was kept short ($\sim
20$~minutes).  In consequence, a long horizontal acrylic piston
(183\,cm~$\times$~7.62\,cm OD, wall thickness 6.35\,mm) was used, exposing a
large surface area of fluid to the vacuum and permitting visual monitoring
during the degassing and injection steps.

Under the injection pressure, the umbilical is expanded against the
constraining copper pipes by the silicone.
The remaining time until the silicone stops flowing freely allows this pressure
to be relieved (the silicone is permitted to escape from both ends).
The finished umbilical can be easily removed from the copper pipes once the
RTV615 is fully cured two days later.

\section{Laser system}

\begin{figure}
  \begin{center}
    \includegraphics[width=7.9in,angle=90]{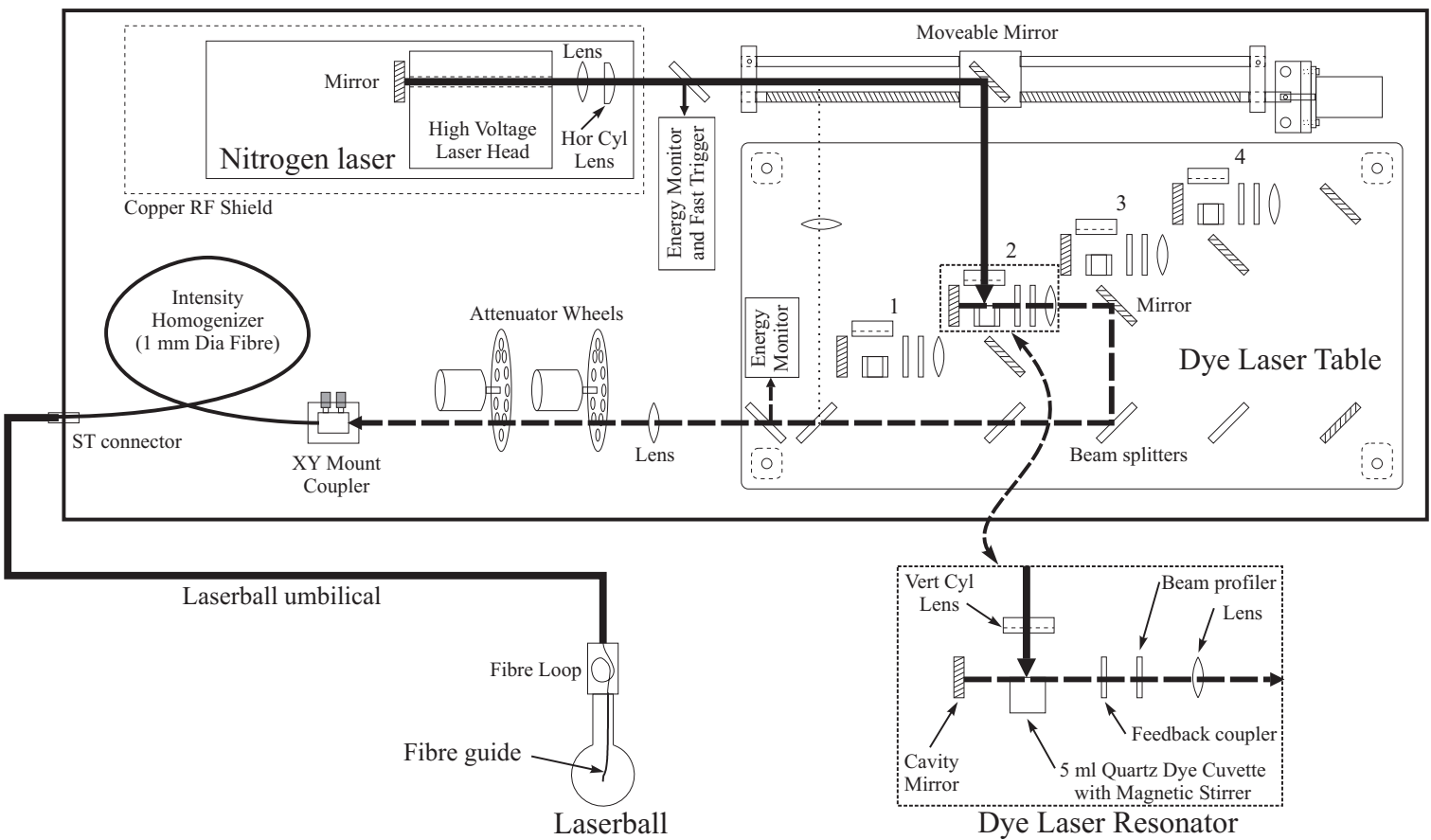}
  \end{center}
  \caption{N$_2$/dye laser system overview.
  Each of the attenuator wheels has 8 positions: open, beam stop, and six
  filters of neutral densities: 1, 2, 3, 4, 5, 6, and 0.1, 0.3, 0.4, 0.6, 0.7, 1.0.
  The four dye laser resonators are illuminated by the N$_2$ laser with
  assistance of a moveable mirror. 
  The outer box represents the heavy aluminum box cover.
  }
  \label{fig:laser_optic_system}
\end{figure}

The laser system consists of a short pulse-length nitrogen laser ($\lambda = 337.1$\,nm)
which can be made to pump a series of up to five laser dyes at longer wavelengths
($\lambda$ = 369, 385, 420, 505 and 619\,nm).
Figure~\ref{fig:laser_optic_system} shows the layout of the laser system,
consisting of the following components:
\begin{enumerate}
\item Commercial N$_2$ TEA (Transversely Excited Atmosphere)
      thyratron triggered ultraviolet pump laser (Class IIIb
      \citep{laserphotonics:ln203c}).

      The laser head is a channel 10\,cm long with a 4\,mm electrode gap and
      a feedback mirror at one end of a high voltage (15\,kV) parallel plate
      capacitor.
      The output is super-radiant and therefore lacks the coherence of a
      cavity mode laser.
      The laser is enclosed in a copper radio-frequency (RF) shield, which
      attenuates the high levels of RF radiation emitted during the rapid high
      voltage discharge across the gas cell.
      This is critical because the RF noise is synchronous with the laser
      light.
\item Optical table where four dye laser resonator units and associated beam
      optics are mounted. 
      The dye laser cell is selected by moving a mirror mounted 
      on a computer controlled lead screw carriage. The fifth dye cell is
      swapped manually.
      
      The nitrogen laser beam is focused through a cylindrical lens to a line
      just inside the dye cell cuvette wall.  The cavity mirror and feedback
      coupler generate a well defined beam along the focused edge of the cuvette
      (mounted at an angle so that feedback from internal reflections are not
      amplified).
      The resulting beam is approximately 2\,mm in diameter with a half-angle
      divergence of 3\,mrad and rectangular diffraction fringes.
      The cuvette holders incorporate a motor that magnetically drives a stirring
      agitator in the bottom of the cuvette.

      The beam optics are set so that the beams from the four dye laser units
      converge on a single axis using four semi-reflective mirrors.
      By putting the longest wavelength in the fourth dye cell and the
      shortest in the first, these steering mirrors partially compensate for
      the higher attenuation of shorter wavelengths in the optical fibres.

\item Two computer controlled attenuator wheels with 8 positions each: open,
      beam stop, 6 coarse adjustment neutral density (ND) filters in the first
      wheel, and 6 finely spaced ND filters in the second wheel. The combined
      neutral density is adjusted to produce pulses with an intensity
      equivalent on average to single photoelectron illumination of all the
      tubes (as mentioned before, this allows about 5\% of all the PMTs to be
      illuminated by each pulse).
\item An intensity homogenizer (1\,mm diameter fibre, 1\,m long) to remove
      pulse-to-pulse beam pattern instabilities.
\item Remote and local system power control and trigger control of pump laser
      and trigger lockout.
\item Pump beam and dye beam laser energy monitors, amplifiers, and a fast
      beam activated event trigger.
      The latter is generated by a fast reverse biased MRD500 PIN diode, that
      produces a negative pulse of 0.8\,ns full width.
      This provides superior timing compared to command triggering of the laser,
      which would include about 5\,ns timing jitter due to the logic circuits
      and the thyratron switch.
\item Heavy aluminum box consisting of a $1.27 \times 61 \times 186$\,cm base
      plate, an angle frame and 1.5\,mm thick sheet aluminum covers on all
      sides.  This box provides mechanical stability for the optical
      components and prevents dust from getting in, and RF or laser
      radiation from getting out.
\end{enumerate}

\begin{figure}
  \begin{center}
    \includegraphics[width=5.4in]{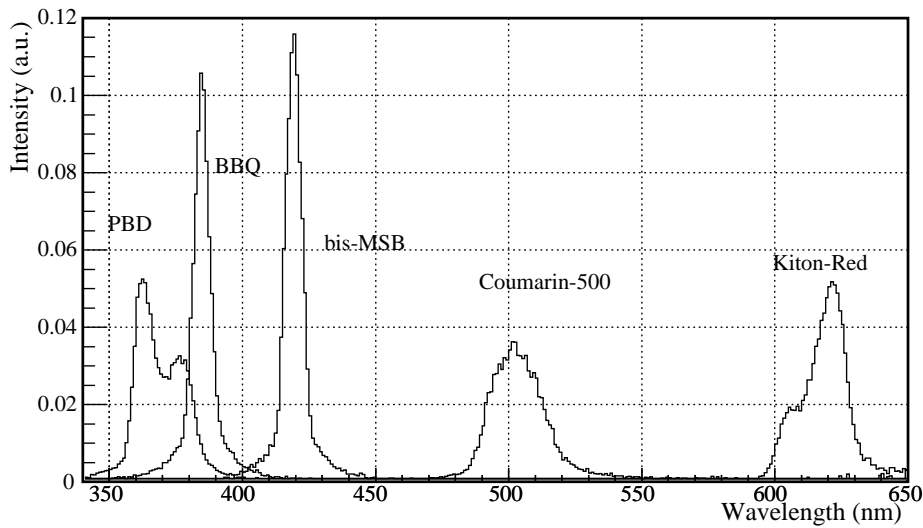}
  \end{center}
  \caption{Wavelength spectra of the laser dyes.}
  \label{fig:dyespectra}
\end{figure}

The dye wavelengths used are listed with other key parameters of the laser
system in Table~\ref{tabl:laser_char}. Figure \ref{fig:dyespectra} shows the
stimulated emission spectra of each of the dyes. The spectra were measured by
using the dyes in a commercial N$_2$/dye laser
unit \citep{photontechnology:gl3300} coupled to a 
spectrophotometer that has a 2~nm resolution. The measured
peak positions and widths are compatible with the manufacturer's nominal values.
Differences in peak wavelength of about 1\% or lower are observed when 
comparing the spectra of dyes used for 3 years in SNO with those of 
newly mixed dyes. Such differences can cause a negligible systematic effect of
about 0.1\% in the refractive index, so the impact of dye degradation in the analysis of
SNO data is not significant.

Table~\ref{tabl:lasersystem_timing}
shows the relevant contributions to timing dispersion in the laser system
coupled through optical fibres (total length 45\,m) to the 
laserball. The fibre dispersion measurements \citep{Ford:98} were made by using a 
setup in which the fibre is looped over a large drum and a small PMT detects 
photons that scatter out of the fibre for each pass of a light pulse produced by 
a fast N$_2$ pulser lamp. The PMT timing spectrum contains multiple peaks offset 
by the loop transit time and the width of those gives the dispersion as a
function of the distance.

\begin{table}
~\bigskip
~\bigskip
\begin{tabular}{ll} \hline
Pulse width  & $t_{\mbox{\sz N$_2$}} \simeq 0.6$~ns \\
             & $t_{\mbox{\sz dyes}} \simeq 0.3$ to 0.5~ns \\ \hline
Pulse Energy & E$_{\mbox{\sz N$_2$}} \sim 100$~$\mu$J/pulse
               or $1.7\times 10^{14}$~photons \\
             & E$_{\mbox{\sz dyes}} \sim 10$ to 30~$\mu$J/pulse \\ \hline
Pulse Rate   & 1 to 45 Hz \\ \hline
\hline
337.1\,nm & N$_2$ laser fundamental, $\sigma_\lambda \simeq 0.1$ nm \\
369\,nm & PBD, $\sigma_\lambda \simeq 10$ nm \\
385\,nm & BBQ, $\sigma_\lambda \simeq 8$ nm \\
420\,nm & Bis-MSB, $\sigma_\lambda \simeq 8$ nm \\
505\,nm & Coumarin 500, $\sigma_\lambda \simeq 14$ nm \\
619\,nm & Kiton Red, $\sigma_\lambda \simeq 10$ nm \\ \hline
\end{tabular}
\caption{Nitrogen/dye laser characteristics for the Laser Photonics LN-203C
  with Exciton dyes.
  The mean wavelengths and approximate RMS widths of the emission spectra (denoted
   by $\sigma_\lambda$) are obtained from the measurements shown in Fig.\ref{fig:dyespectra}.
  }
\label{tabl:laser_char}
~\bigskip
\end{table}


\begin{table}
~\bigskip
  \begin{tabular}{llll|l}\hline
    Wavelength & Laser width & Fibre optic & Laserball  & Net width \\
               &             & dispersion  & dispersion &           \\
    \hline
    337.1 nm & 0.6\,ns        & 0.5\,ns & 0.3\,ns & 0.8\,ns \\
    369 nm   & 0.3 -- 0.5\,ns & 0.5\,ns & 0.3\,ns & 0.7 -- 0.8\,ns \\
    \hline
  \end{tabular}
  \caption{Laser system timing characteristics for 45~m of optical fibre.}
  \label{tabl:lasersystem_timing}
  ~\bigskip
\end{table}

All the source manipulation operations can be done remotely, except for the
actual source insertion (and removal) within the AV. 
A fifth resonator unit was recently added to the optical table, so all
wavelengths can be selected with no manual intervention. Therefore, with the
exception of the occasional replacement of the laser N$_2$ supply, the optical
calibration system can be operated remotely, 
which makes it easier to conduct around-the-clock, multiple-day calibrations since 
access to the SNO underground laboratory is limited.

\section{Performance of the optical source}
\label{sect:performance}

The laserball (LB) distribution is implemented in the OCA model as a combination of a
a 2-dimensional $12\times36$ matrix of the polar angle cosine
$\cos\theta_{\rm{LB}}$ versus the azimuthal angle $\phi_{\rm{LB}}$ and a 
sixth-order polynomial function on $\cos\theta_{\rm{LB}}$ only.
The polynomial function allows the model to accomodate a rapidly varying
distribution without increasing the size of the matrix.

\begin{figure}
  \begin{center}
    \includegraphics[width=5.4in]{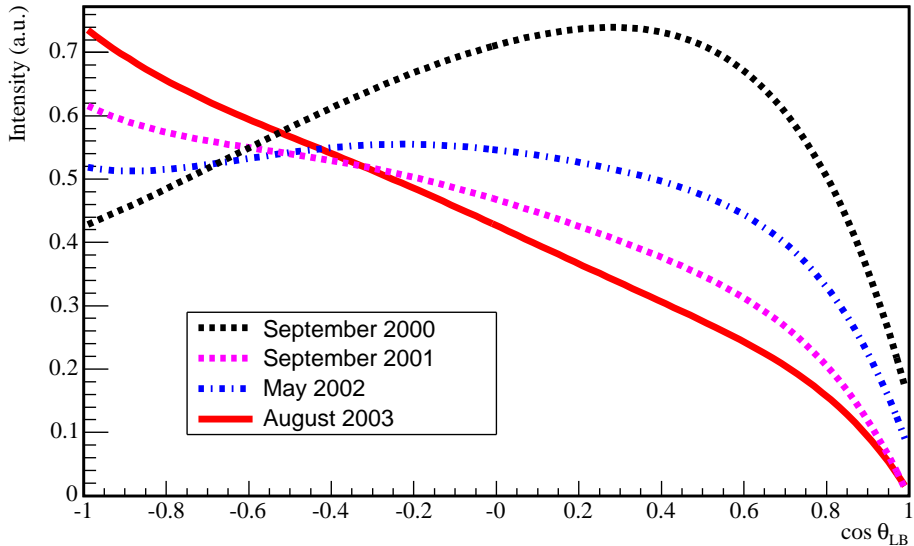}
  \end{center}
  \caption{Light intensity of the laserball as a function of the polar angle 
   cosine at 505 nm, determined by the OCA. The direction vertically upwards 
   from the laserball corresponds to $\cos\theta_{\rm{LB}}=1$.
    }
  \label{fig:fit_mask}
 \end{figure}

Figure~\ref{fig:fit_mask} shows the far-field laserball intensity
distribution, determined by the OCA for four different calibration scans at 
505~nm, as a function of the polar angle cosine.
A strong variation of the polar light distribution is expected at
$\cos\theta_{\rm{LB}}$ close to 1 due to shadowing by the source support
hardware.
Away from the top region, the polar variation of the light intensity can be up
to $\pm 40$\% and is likely
determined at assembly by the exact configuration of gaps between the fibre
bundle and the rigid fibre (since this can influence the angular distribution of
light arriving to the gel). 
The laserball used in the September\,2000 scan is different from the one used in
the later scans. The variations between the September\,2001, May\,2002 and
August\,2003 are likely due to variations in the laserball-umbilical coupling
and degradation of the gel.

\begin{figure}
  \begin{center}
    \includegraphics[height=6in]{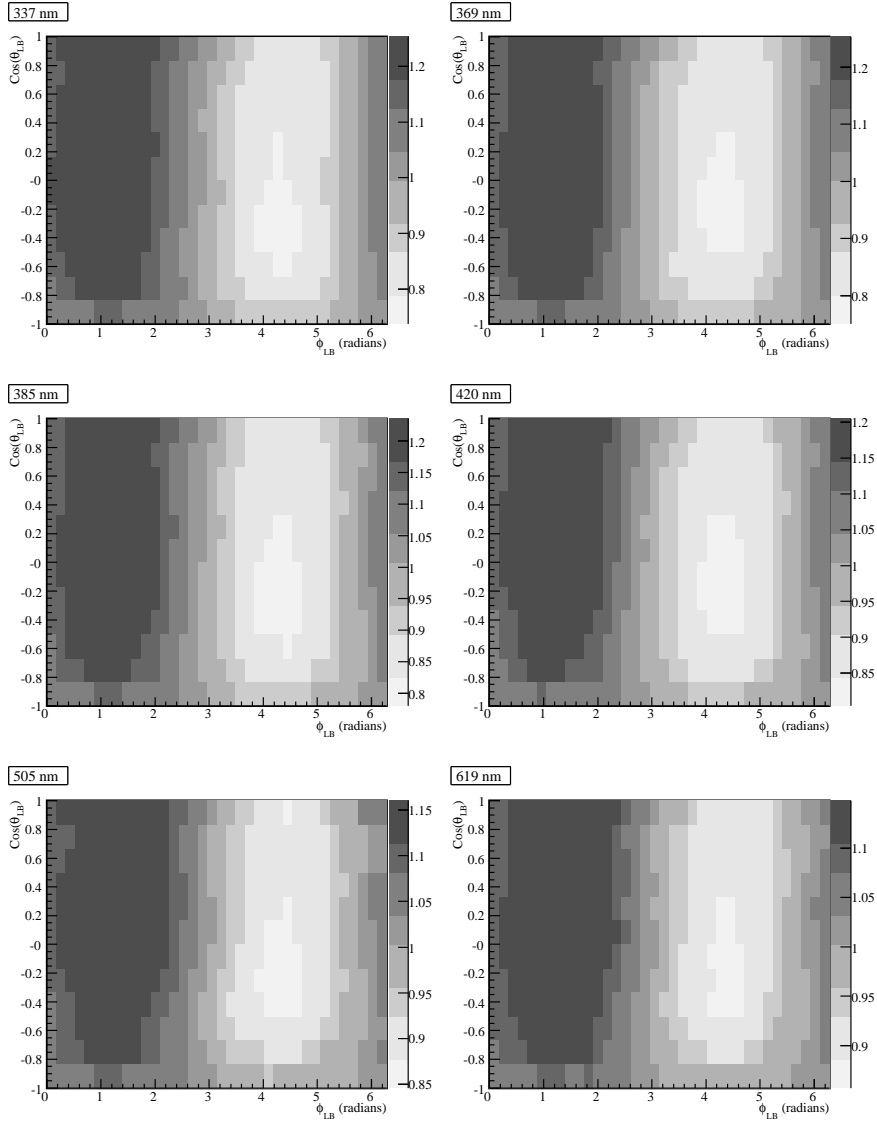}
  \end{center}
  \caption{Relative light intensity of the laserball in function of polar
  (residual only) and azimuthal angles, determined by the OCA for data taken
    in September 2000. 
	The direction vertically upwards from the laserball
    corresponds to $\cos\theta_{\rm{LB}}=1$.
    }
  \label{fig:fit_laserdist_sept}
 \end{figure}

Figure~\ref{fig:fit_laserdist_sept} shows the laserball 
intensity distribution determined by the OCA for each wavelength 
\citep{Moffat:01}, after taking into account the polar angle dependence shown in 
Figure~\ref{fig:fit_mask}.  The grey scale indicates the relative light 
intensity as a function of
the cosine of the polar angle, $\cos\theta_{\rm{LB}}$, and azimuthal angle,
$\phi_{\rm{LB}}$.
The light intensity varies by $\pm 15$\% at 619\,nm, increasing to $\pm 25$\%
at 337.1\,nm 
The azimuthal light distribution is dominated by a dipole distribution, 
caused by misalignment of the fibre tip inside the laserball.  The
larger asymmetry at shorter wavelengths is likely the effect of increased
absorption in the laserball gel and glass spheres. 

The OCA must determine average
optical properties of the whole detector. So it is 
important for the variation in azimuthal distribution to be smooth and
relatively small, as 
shown in Figure~\ref{fig:fit_laserdist_sept}, to guarantee a good sampling of 
all the PMTs.
Fluctuations in the laserball intensity that occur in a short timescale (less than
a week) can affect the accuracy of the OCA
during a multi-day scan. Hence, runs are repeated at specific positions
throughout the scan to ensure the stability of the light intensity.

\section{Conclusions}

An accurate calibration of the optical properties of SNO is a vital step in
understanding the performance of the detector.  Measurements with the
laserball system at six wavelengths from 337.1\,nm to 619\,nm allow the optical
properties to be sampled over the detectable Cherenkov light spectrum.  These
optical data for the OCA are taken approximately twice per year, each time for
one week
around the clock.  The system is fully remotely controlled so that continuous
underground access to the SNO laboratory is not necessary.

The short pulse length and good uniformity characteristics of the light
distribution permit the simultaneous relative timing and collected charge
calibrations(PCA) of all 9456~PMT channels to be accomplished efficiently.  The
timing calibration is performed at 505\,nm because the light intensity varies by
only $\pm$15\%, and absorption and scattering are low in all three media (\heavywater,
\lightwater\ and acrylic).  These calibrations are done monthly.

The laserball system has proven very robust over more than 2500 hours of operation.
Limited maintenance has been required, such as new charge transfer boards
for the laser head and realignment of the mirrors.  The umbilical cables, made
using a novel manufacturing technique, are flexible, clean, robust and
versatile: the basic umbilical design was easily modified to provide gas,
electrical connections or simply act as a support line for each of the eight 
other SNO calibration devices.

The optical calibration hardware enables the detector optics to be
probed \emph{in situ} and provides the essential timing reference for all PMT
channels.  These calibrations are crucial to the energy and position
reconstruction of events, and are essential to the precision of SNO's solar
neutrino data analysis.


\section*{Acknowledgements}

This research was supported in part by the Natural Sciences and Engineering
Research Council of Canada, and the Fonds pour les Chercheurs et l'Aide \`a la
Recherche of the province of Qu\'ebec, Canada.


\end{document}